\documentclass[preprint,a4paper]{aastex63}

\usepackage{amsmath,amssymb}
\usepackage{natbib}
\usepackage{graphicx}
\usepackage{color}
\usepackage{url}
\usepackage{upgreek}

\usepackage[varg]{txfonts}
\usepackage{gensymb}
\usepackage{multirow}

\usepackage{appendix}
\usepackage{threeparttable}

\usepackage{tikz,hyperref}

\received{June 9, 2020}
\accepted{October 6, 2020}
\submitjournal{Astronomy \& Astrophysics}

\shorttitle{A nonlinear solar magnetic field calibration method}
\shortauthors{J. Guo et al.}

\begin{document}

\title{A nonlinear solar magnetic field calibration method for the filter-based magnetograph by the residual network}

\correspondingauthor{Kaifan Ji}
\email{jkf@ynao.ac.cn}

\author[0000-0003-4768-422X]{Jingjing Guo}
\affiliation{Key Laboratory of Solar Activity, National Astronomical Observatories, Chinese Academy of Sciences, Beijing 100101, China}
\affiliation{University of Chinese Academy of Sciences, Beijing 100049, China}

\author[0000-0003-2686-9153]{Xianyong Bai}
\affiliation{Key Laboratory of Solar Activity, National Astronomical Observatories, Chinese Academy of Sciences, Beijing 100101, China}
\affiliation{University of Chinese Academy of Sciences, Beijing 100049, China}

\author[0000-0003-2714-6811]{Hui Liu}
\affiliation{Yunnan Observatories, Chinese Academy of Sciences, Kunming 650216, China}

\author[0000-0002-3238-0779]{Xu Yang}
\affiliation{Big Bear Solar Observatory, 40386 North Shore Lane Big Bear City, CA 92314-9672, U.S.A.}

\author[0000-0003-1988-4574]{Yuanyong Deng}
\affiliation{Key Laboratory of Solar Activity, National Astronomical Observatories, Chinese Academy of Sciences, Beijing 100101, China}
\affiliation{University of Chinese Academy of Sciences, Beijing 100049, China}

\author[0000-0001-9489-0101]{Jiaben Lin}
\affiliation{Key Laboratory of Solar Activity, National Astronomical Observatories, Chinese Academy of Sciences, Beijing 100101, China}

\author[0000-0002-5152-7318]{Jiangtao Su}
\affiliation{Key Laboratory of Solar Activity, National Astronomical Observatories, Chinese Academy of Sciences, Beijing 100101, China}
\affiliation{University of Chinese Academy of Sciences, Beijing 100049, China}

\author[0000-0003-1675-1995]{Xiao Yang}
\affiliation{Key Laboratory of Solar Activity, National Astronomical Observatories, Chinese Academy of Sciences, Beijing 100101, China}

\author[0000-0001-8950-3875]{Kaifan Ji}
\affiliation{Yunnan Observatories, Chinese Academy of Sciences, Kunming 650216, China}

\begin{abstract}

The method of solar magnetic field calibration for the filter-based magnetograph is normally the linear calibration method under weak-field approximation that cannot generate the strong magnetic field region well due to the magnetic saturation effect. We try to provide a new method to carry out the nonlinear magnetic calibration with the help of neural networks to obtain more accurate magnetic fields. We employed the data from \textit{Hinode}/SP to construct a training, validation and test dataset. The narrow-band Stokes I, Q, U, and V maps at one wavelength point were selected from all the 112 wavelength points observed by SP so as to simulate the single-wavelength observations of the filter-based magnetograph. We used the residual network to model the nonlinear relationship between the Stokes maps and the vector magnetic fields. After an extensive performance analysis, it is found that the trained models could infer the longitudinal magnetic flux density, the transverse magnetic flux density, and the azimuth angle from the narrow-band Stokes maps with a precision comparable to the inversion results using 112 wavelength points. Moreover, the maps that were produced are much cleaner than the inversion results. The method can effectively overcome the magnetic saturation effect and infer the strong magnetic region much better than the linear calibration method. The residual errors of test samples to standard data are mostly about 50 G for both the longitudinal and transverse magnetic flux density. The values are about 100 G with our previous method of multilayer perceptron, indicating that the new method is more accurate in magnetic calibration.
\end{abstract}

\keywords{Solar Magnetic fields, Polarized observation; Convolution neural network, Residual network}

\section{Introduction}
\label{sect:intro}

 \begin{figure*}[htbp]
	\centering
	\includegraphics[width=.8\linewidth]{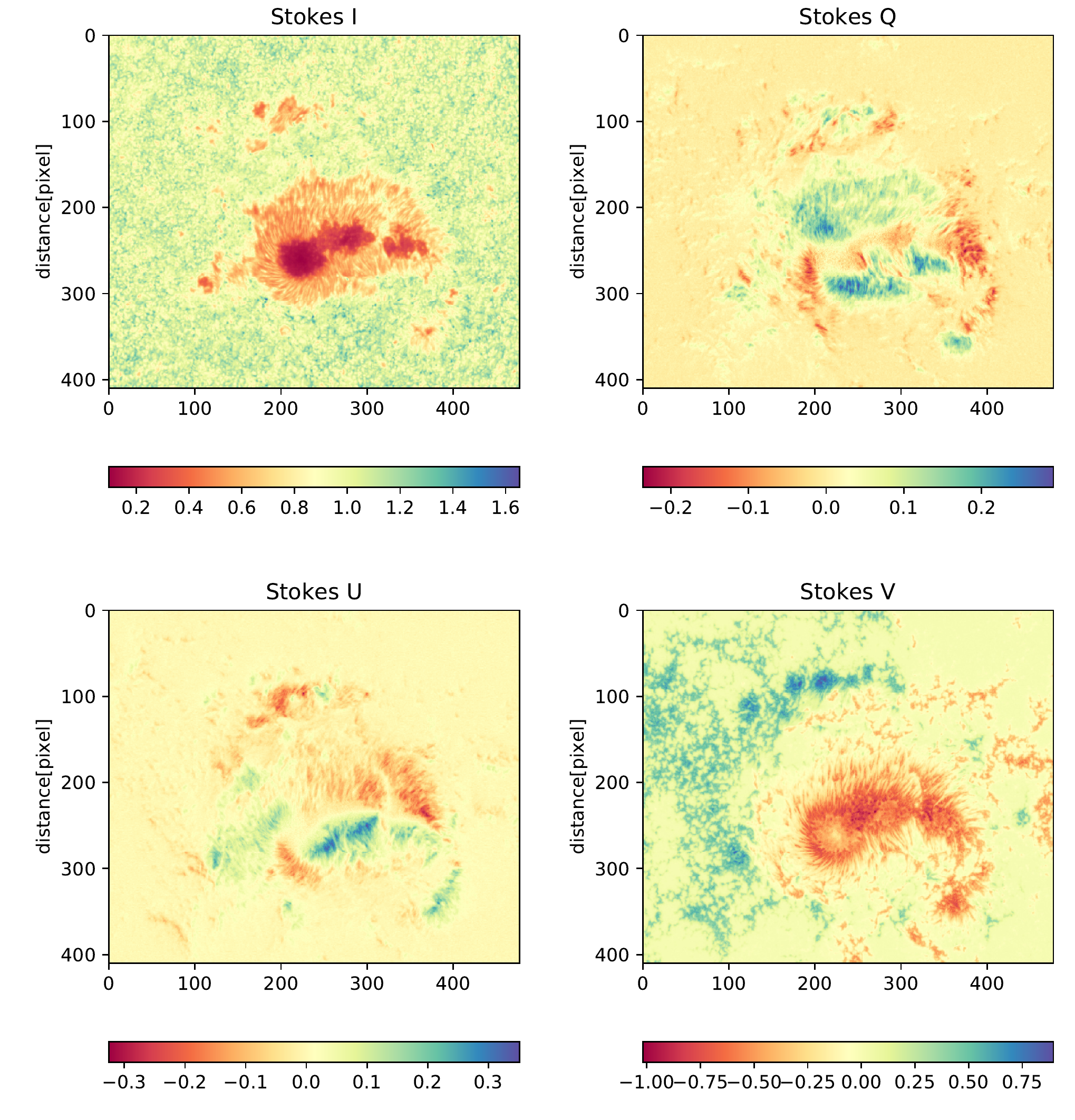}
	\caption{Stokes $ I $, $ Q $, $ U $, and $ V $ at -0.063 \AA\ apart from the line center of Fe \textsc{I} 6301 \AA\ as the input parameters. The data were observed at 10:47 UT on 2014 September 11 in NOAA AR 12158. These maps of Stokes parameters were normalized quantities by Equation 6 after data preprocessing.}
\end{figure*}

\begin{figure*}[htbp]
	\centering
	\includegraphics[width=1\linewidth]{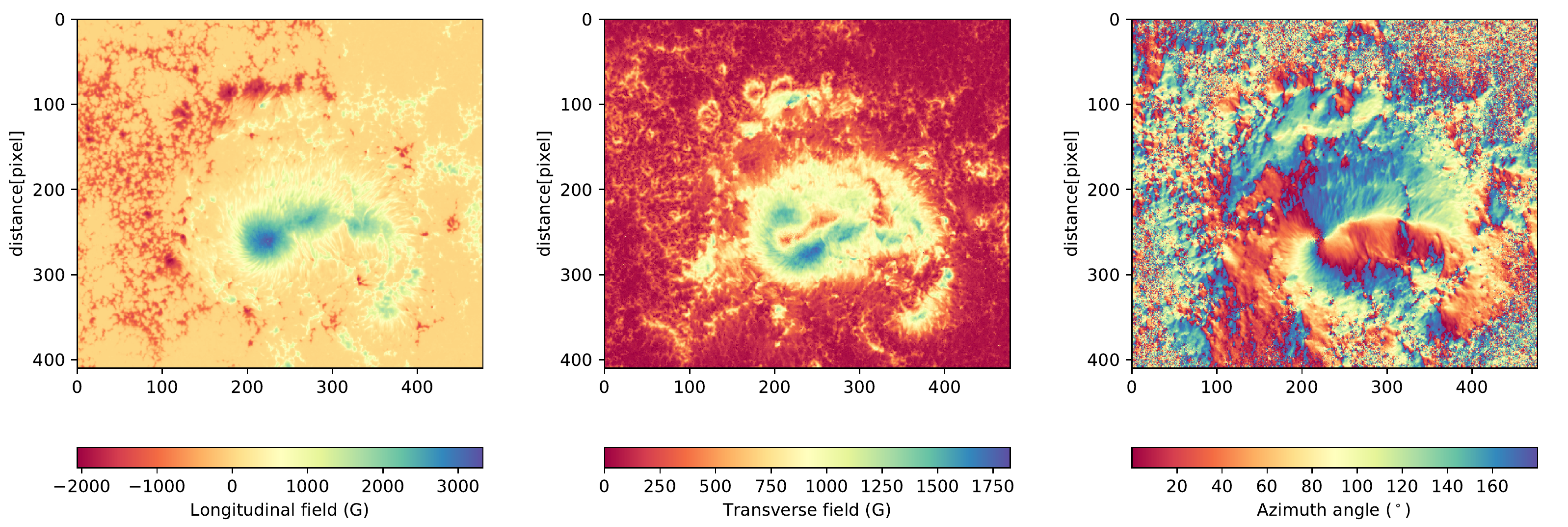}
	\caption{$B_l$ and $B_t$ (the main parameters we are concerned with in our study) without considering $\alpha $ or $ \varphi $ (azimuth angle) as the target parameters.}
\end{figure*}

At present, the solar magnetic field is generally measured indirectly by means of the Zeeman effect \citep{hale1908probable} of the magnetically sensitive solar spectral lines. According to the different focal plane equipment, the optical magnetograph can be classified as the systems of the systems of the grating spectrographs (SGs) and filtergraphs (FGs) \citep{Lin2001solar,2019OptEn..58h2417I}. Additionally, the integral field solutions are being developed to solve the inability of SGs and FGs to cover the desired spatial
and spectral field of views simultaneously \citep{2019OptEn..58h2417I}.

The advantage of the spectral magnetograph is its capability to obtain a primary spectrum, so the magnetic field is generally inferred by inversion techniques which have been the most useful tool since the early 1970s  \citep{1987ApJ...322..473S,1992ApJ...398..375R,2001ASPC..236..487S,2015IAUS..305..225A,2019ApJ...875..127L}. The inversion techniques fit the observed and the synthetic stokes profiles calculated from the solution of the radiative transfer equation considering the appropriate solar atmospheric parameters, such as the magnetic field and temperature, for example \citep{2016LRSP...13....4D}. The method estimates a highly accurate magnetic field at the expense of a rather large computing resource consumption, since the Stokes inversion of each pixel takes substantial calculating time. Benefiting from the improved computer capabilities, deep learning has greatly developed in recent decades \citep{goodfellow2016deep}. Therefore, novel methods and techniques have been tested and acquired some advanced results on magnetic field inversion. Socas-Navarro et al. first utilized a principle component analysis (PCA) to infer the magnetic field strength and inclination angle from Stokes profiles, which is extremely fast and stable \citep{2000A&A...355..759R,2001ApJ...553..949S,2001ASPC..236..487S}. Meanwhile, T. A. Carroll and J. Staude applied artificial neural networks (ANNs) to train the network between Stokes profiles and solar atmospheric parameters and retrieved a good estimation for the magnetic field vector \citep{carroll2001inversion}. The statistical machine-learning techniques based on Mercer's kernel were proposed to inverse the vector magnetic field by \citet{teng2015application}. Convolutional neural networks (CNN) were applied to Stokes inversion by \citet{2019A&A...626A.102A}, \citet{2020arXiv200503945L}, and \citet{2020arXiv200602005M}. Those methods mostly employed the numerical Magneto-Hydro-Dynamical (MHD) simulated data in training as well as during the validating process, and all of them worked on multi-wavelength polarimetric signals. 

The traditional filter-based magnetograph commonly employs a single wavelength point in regular magnetic field observations to obtain high temporal resolution. One way to carry out magnetic calibration is via the linear calibration method based on a weak-field approximation by comparing results with the Stokes inversion from multi-wavelength points \citep{bai2013calibration,2014MNRAS.445...49B,2004ChJAA...4..365S}. The purpose of magnetic calibration is to obtain the appropriate coefficients between Stokes parameters and magnetic vectors,  that is, the magnetic flux density in the longitudinal direction ($B_{l}$), the transversal direction ($B_{t}$), and $ \varphi $. For a filter-based magnetograph, the previously mentioned coefficients can be obtained by Equations 1, 2, and 3 under the assumption of a weak field \citep{Stenflo1994}. We note that $ C_{l} $ and $ C_{t} $ are the linear calibration coefficients for $B_{l}$ and $B_{t}$, respectively, that is:

\begin{equation}
B_{l}=C_{l}V   
\end{equation}
\begin{equation}
B_{t}=C_{t} (Q^{2}+U^{2})^{1/4}
\end{equation}
\begin{equation}
\varphi=0.5\arctan(U/Q)
.\end{equation}
For the strong magnetic field region, the weak-field approximation is not satisfied due to the magnetic saturation effect. In this case, the linear calibration method has a large error when recovering the strong magnetic field. The \textit{Narrowband Filter Imager} onboard \textit{Hinode} \citep{tsuneta2008solar} fits the respective linear calibration coefficients on the different solar structure so as to avoid magnetic saturation in the strong-magnetic field region.\citep{chae2007initial}. Prior calibration studies \citep{Guo1989SOLAR,2007ChJAA...7..281Z} for the \textit{Solar Magnetic Field Telescope} (SMFT) installed at the Huairou Solar Observing Station (HSOS) also verified this conclusion. With ANNs \citep{carroll2001inversion,2003NN.....16..355S,2005ApJ...621..545S} and deep learning (DL), which is generally utilized  \citep{2018A&A...614A...5D,2019NatAs...3..397K,2019AdAst2019E..29Z,Liuhui2019,2020ApJ...891L...4P}, \citet{Guo_2020} attempted to carry out magnetic field calibration for a filter-based magnetograph using multi-layers perceptron (MLP) without considering the spacial correlation to build models to infer magnetic fields. In this paper, we try to use the state-of-art CNN method, considering spacial correlation, to calibrate the magnetic field for a filter-based magnetograph.

The Full-disk MagnetoGraph (FMG) \citep{2019RAA....19..157D}, which is a filter-based magnetograph working on Fe \textsc{I} 532.42 nm onboard the \textit{Advanced Space-based Solar Observatory} (ASO-S) \citep{2019RAA....19..156G}, is being designed by National Astronomical Observatories, Chinese Academy of Sciences and is scheduled to be launched around 2022 to measure full-disk vector magnetic fields. The preliminary method for data reduction and calibration of the FMG has been studied by \citet{2019RAA....19..161S}. So this work is helpful for the nonlinear magnetic field calibration of FMG. 

The paper is organized as follows. In Section 2, the process of obtaining a dataset is introduced. In Section 3, we describe the neural network structure and training strategies. In Section 4, we evaluate the accuracy and performance of the trained models using the test dataset. Finally, the conclusions are given in Section 5 with a discussion on the properties and the future work. 

\section{Datasets}
\label{sec:data}

The purpose of the study is to improve the quality and accuracy of actually observing magnetic field calibration for a filter-based magnetograph working at a single wavelength point. The dataset is selected from Hinode/SP \citep{tsuneta2008solar} and the detailed data acquisition process is described in \citet{Guo_2020}. For a better comparison with the MagMLPs, we employed the same datasets to train the CNN models. The datasets contain 176 frames of active regions including level 1 data and level 2 data \citep{2013SoPh..283..601L}. Level 1 data include the Stokes I($\lambda$), Q($\lambda$), U($\lambda$), and V($\lambda$) polarized maps with 112 wavelength points. Level 2 data are the magnetic parameters coming from the Stokes inversion of the High Altitude Observatory (HAO) Merlin inversion code, which was developed by the Community Spectra-polarimetric Analysis Center. A brief description of the training set, validation set, and test set is listed in Table 1. The wavelength point was taken from the spectral polarimeter profiles at -0.063 \AA\ apart from the line center of Fe \textsc{I} 6301 \AA\ to simulate the observation of the filter-based magnetograph. That is because the point at this wavelength has the relatively large value on the first derivative of  the spectral profile and it is the most sensitive point to magnetic signals. Furthermore, regarding the actual observation of the filter-based magnetograph, the observation wavelength is usually chosen according to this principle. Then the Stokes I, Q, U, and V maps at this wavelength are considered as the input parameters. A sample of an active region in NOAA AR 12158 in the training set is presented in Figure 1.

From the level 2 data, we can directly obtain the magnetic field ($ B $) by taking the filling factor ($ \alpha $), the inclination angle ($ \theta $), and the azimuth angle ($\varphi$) into account. For a filter-based magnetograph, it is not suitable for the diagnosis of a filling factor ($ \alpha $) due to the limited spectral information. For example, the filling factor is also not included in the data products for both HMI \citep{2011SoPh..273..267B} and SO/PHI \citep{2019AGUFMSH21D3292M} data. So we used the longitudinal magnetic flux density ($ B_{l} $) and transverse magnetic flux density ($ B_{t} $) without considering the $ \alpha $ or $ \varphi $ as the target parameters. The corresponding $ B_{l} $ and $ B_{t} $ can then be obtained by Equations 4 and 5:\begin{equation}B_{l}=\alpha B \cos(\theta),\end{equation}
\begin{equation}B_{t}=\alpha B \sin(\theta ).\end{equation} \footnote{On the code's webpage \url{https://www2.hao.ucar.edu/csac/csac-data/sp-data-description}, we found that "The \textit{Hinode/SP} inversions solve for the fill factor $ \alpha $, the observed Stokes \textit{I} profile $ I_{obs} $ is fitted with $ \alpha*I_{mag} +(1-\alpha )*I_{scatt} $, where $ I_{mag} $ is the magnetized component and $ I_{scatt} $ is the scattered light profile." Here we do not follow this concept strictly. Please be careful if you use our method; we will elaborate on this in the appendix.} The sample of them in NOAA AR 12158 inferred by inversion code are shown in Figure 2.

\begin{table}[]
	\caption{List for the training set, validation set, and test set.}
	\begin{threeparttable}              
		\begin{tabular}{lllll}
			\hline
			\multicolumn{1}{c}{\begin{tabular}[c]{@{}c@{}}DATA\\  SET\end{tabular}} & 
			\multicolumn{1}{c}{\begin{tabular}[c]{@{}c@{}}DATE\\ (YYYYMMDD)\end{tabular}}                                                             & \multicolumn{1}{c}{NO.}                                   & 
			\multicolumn{1}{c}{TOTAL} & 
			\multicolumn{1}{c}{RATIO} \\ \hline
			\begin{tabular}[c]{@{}l@{}} Training\\  Set \end{tabular}                 & 
			\begin{tabular}[c]{@{}l@{}}20140726-20141220\\ 20150101-20150216\\  20170905-20170907\end{tabular}                                                             & 
			\begin{tabular}[c]{@{}l@{}}98\\ 14\\9\end{tabular}            & 121                                                                       & 0.69                      \\ \hline
			\begin{tabular}[c]{@{}l@{}} Validate\\ Set \end{tabular}                 &
			\begin{tabular}[c]{@{}l@{}} 20170822-20170904 \\ 20170907-20170929\end{tabular}                                                             & 
			\begin{tabular}[c]{@{}l@{}} 9 \\9 \end{tabular}                                                       & 18                                                                        & 0.10                      \\ \hline
			\begin{tabular}[c]{@{}l@{}} Test\\  Set \end{tabular}                     & \begin{tabular}[c]{@{}l@{}}20110728-20111107\\ 20140104-20140418\\ 20141021-20141024\\ 20180206-20180621\\ 20190321-20190508\end{tabular} & \begin{tabular}[c]{@{}l@{}}9\\ 12\\ 4\\ 5\\ 7\end{tabular} & 37                                                                        & 0.21                      \\ \hline
		\end{tabular}
		\begin{tablenotes}
			\footnotesize
			\item[1]  All the Hinode/SP data are provided by the Community Spectropolarimetric Analysis Center (http://www2.hao.ucar.edu/csac).
			
		\end{tablenotes}
	\end{threeparttable}
\end{table}

On the basis of analysis for influencing factors, the pre-processing to input and target data set includes data cleaning and unifying the input data. Certain data processing methods have been applied to select a strong magnetic field region and improve the signal-to-noise (S/N). The Stokes $ I $ intensity was normalized by dividing by the median value of the quiet region itself. The components $ Q $, $ U $, and $ V $ were indirectly divided by the Stokes $ I $ to unify data. The equations are presented as 
\begin{equation}
I_{\mathrm{norm}} = \frac{I}{I_{\mathrm{median}}},\quad Q_{\mathrm{norm}} = \frac{Q}{I},\quad U_{\mathrm{norm}}=\frac{U}{I},\quad V_{\mathrm{norm}}=\frac{V}{I}.
\end{equation}

In this study, Stokes $ I $, $ Q $, $ U $, and $ V $ were used as input parameters, and $ B_{l} $, $ B_{t} $, $\varphi$ were used as the output parameters. The sizes of each data sample are different. Using maps of different sizes to train the models increases the difficulty of training. So the maps of the training set and validation set are cut into the same shape to train. If the maps are cut too small, the features of a map decrease. If the maps are cut too large, this increases the training requiring memory of a graphics card. As we already know, the training of neural networks need a large amount of samples. For one epoch of the CNN training, we randomly selected a block of 65 pixel $ \times $ 65 pixel from each of the total 121 sample images. We repeated this kind of selection and training process for tens of thousands of epochs to augment the number of training samples. And then we did the same for geometric transformations of digital images by rotating the blocks by 0\degree, 90\degree, 180\degree, or 270\degree or, alternatively, by mirroring the blocks to further increase the number of training set. 

\section{Method}
\label{sec:method}
In our previous work, the multilayer perceptron is a pixel to pixel neural network, without considering the spatial correlation of images. In this paper, we utilize the CNN \citep{Le1989Cun} architecture to build the neural network model. The residual network (ResNet) \citep{He2016Deep} is one of the famous CNN architectures, which was first developed and published in 2016. Its major improvement is reformulating the layers as learning residual functions, thus making it easier to optimize structures and gain accuracy from a considerably increased depth. The similar architecture has been used by many authors (e.g.,  \citet{2017A&A...604A..11A,2019A&A...626A.102A}) with fantastic performance. Based on ResNet, the CNN models for magnetic calibration are built and then trained. In this case, each output parameter has a different unit and S/N and each parameter needs to be set to a different weight in the loss function. We have not found a perfect solution for assigning weights for each output parameter. In order to gain more accurate models, we abided by the scenario that one model is built for each magnetic parameter. 

The design of the structure for a network model is vital for its performance. According to our experiments, residual blocks combined with the basic convolution layer can best calibrate the magnetic field. The structure of our model is shown in Figure 3. Every convolution layer contains many training parameters and generates many feature maps through a convolution filter extracting a feature \citep{726791}. In our model, we set a convolution layer with 64 filter kernels, with a size of 3 pixel $\times$ 3 pixel to produce feature maps for the initial input parameters. Additionally, all of the filter kernels have a size of 3 pixel $\times$ 3 pixel. The kernel size is crucial for model performance for two reasons: A  kernel that is too large would over-smooth the training result, while a kernel that is too small could reduce the spatial correlation. Then, we put the linear relation into a nonlinear one through a rectified linear unit ReLU activation function, which has a good performance and is extensively applied in the DL \citep{Nair2010Rectified}, defined as:
\begin{equation}
\mathrm{ReLU}(x)=\max(0,x)=\left\{
\begin{aligned}
x &, &  x \geqslant 0 , \\
0 &, &  x < 0 .
\end{aligned}
\right.
\end{equation}

We then set a ReLU activation function to provide the nonlinear properties. These steps were done to fuse the input maps by extracting more feature maps before the residual blocks. 

\begin{figure}[!ht]
	\centering
	\includegraphics[width=1\linewidth]{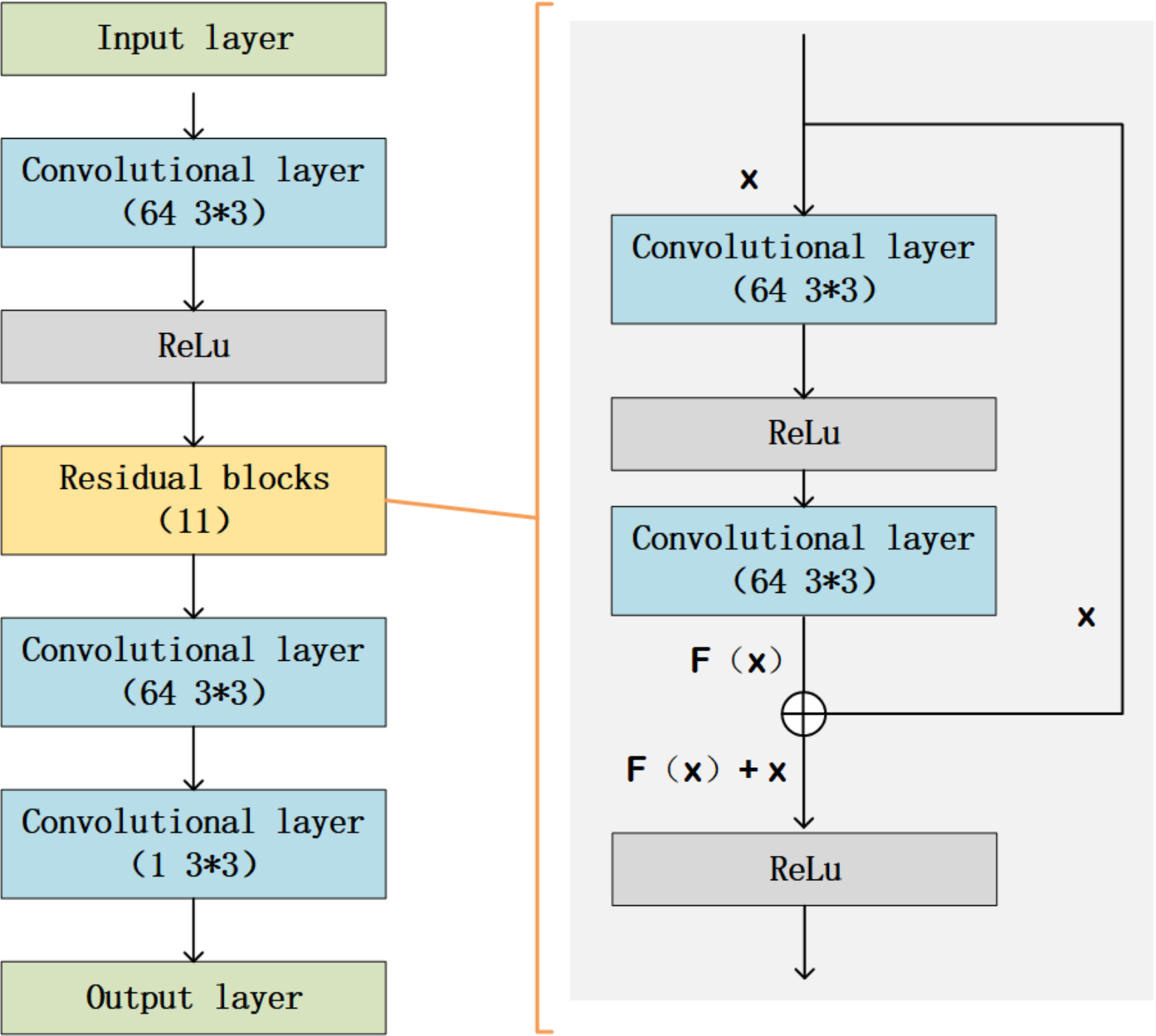}
	\caption{Schematic architecture of the ResNet used in this paper.}
\end{figure}

\begin{figure*}[htbp]
	\centering
	\includegraphics[width=1\linewidth]{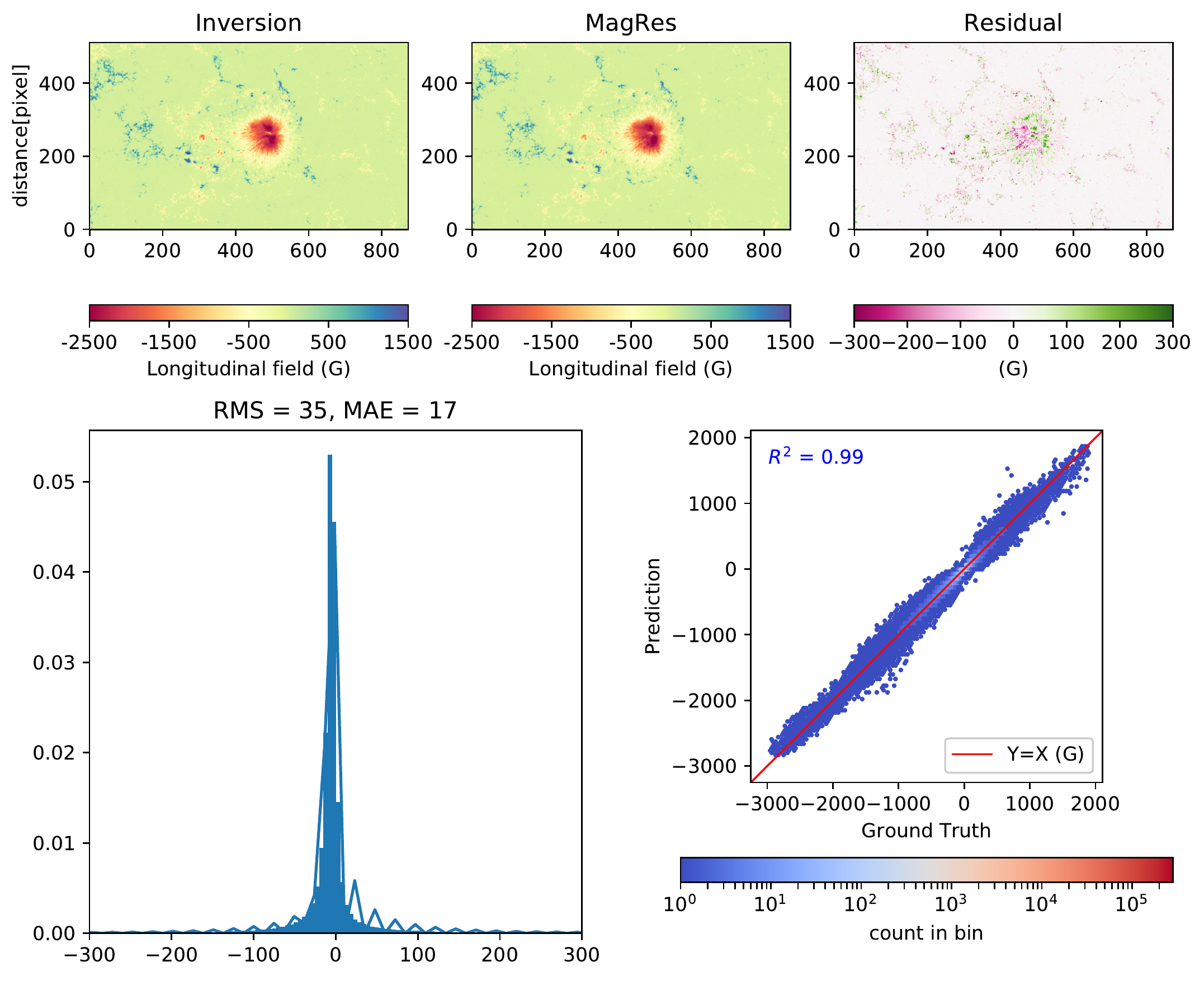}
	\caption{MagRes calibrations for longitudinal magnetic flux density. The upper panels (from left to right) are the inverted result, the MagRes result, and their residual difference, respectively. The lower left panel is the histogram of the residual error with its Gaussian kernel density curve. The lower right panel is the scatter diagram, which identifies the density of the inversion results with the testing results.}
\end{figure*} 

There are 11 residual blocks in this model. Each residual block contains a residual part and an identity of input $ x $. The residual part is made of five functional parts, as is shown in the left panel of Figure 3. There are two convolution layers with the same kernels as the initial convolution layer. A ReLU layer is set as the middle layer to provide nonlinear mapping. The input $ x $ goes through the residual part to get to the residual $ F(x) $. The output of the residual block is the input identity $ x $, which adds the $ F(x) $, then passes a ReLU layer. We named the network model MagRes.

The MagRes network was trained with a training batch size of 28 and a validation batch size of 18. There are about 9.26 thousand free parameters in the network model. It takes, on average, 21 seconds per epoch to train. The mean squares error (MSE) is a loss function, the Adam algorithm \citep{2014arXiv1412.6980K} is an optimizer, and the initial learning rate is $ 1e^{-4} $. The loss function MSE is a common method used in regression networks, which is defined as follows:
\begin{equation}
\mathrm{MSE} = \dfrac{1}{n}\sum_{i=1}^{n}(y_{i}-\hat{y_{i}})^{2}.
\end{equation}
In the measurement of the vector magnetic field, there is an inherent 180 $ \degree $ ambiguity in the field perpendicular to the line-of-sight, as inferred from observations of linear polarization in magnetically sensitive spectral lines \citep{2006SoPh..237..267M}. This effect results in $ \varphi $ being equivalent to $ \varphi + 180\degree $ in the azimuth angle. In using the same training process as other parameter models, the values around 0 $ \degree $ or 180 $ \degree $ are inferred as a discrete distribution in the range [0,180]$ \degree $. The scatter graph of predictive values and target data looks like a reversing N shape. Therefore, we continued to test the converged model to correct it by changing the loss function, as can be seen in Equation 9. This is the case because the azimuth angle is a circular quantity with 180\degree; the angular difference between the predictive value and the target value should not be more than 90\degree. So if $ |y_{i}-\hat{y_{i}}| > 90 $, we used $ (180-|y_{i}-\hat{y_{i}}|) $ instead, thus:

\begin{equation}
\mathrm{MyLoss} = \dfrac{1}{n}\sum_{i=1}^{n}E_{i}^{2}, 
\end{equation}

where
\begin{equation} E_{i} = \left\{
\begin{array}{rcl}
y_{i}-\hat{y_{i}} & & {|y_{i}-\hat{y_{i}}| \leq 90 }\\
180-|y_{i}-\hat{y_{i}}| & & {|y_{i}-\hat{y_{i}}| > 90}
\end{array} \right.. 
\end{equation}

This work uses Keras 2.2.2, which takes the TensorFlow 1.5.0 as computing background, as the modeling environment of deep learning. For accelerating the training process, GPUs are extensively leveraged in DL, especially for the CNN. Our study mainly works on a NVIDIA Quadro P2000 (set up in our Dell T3620 graphic workstation) with 1024 CUDA Cores, large 5GB GDDR5 on-board memory, and a NVIDIA Tesla P100 (provided by Ali cloud) with more than 21 teraFLOPS of a 16-bit floating-point. 

\begin{figure*}[!ht]
	\centering
	\includegraphics[width=1\linewidth]{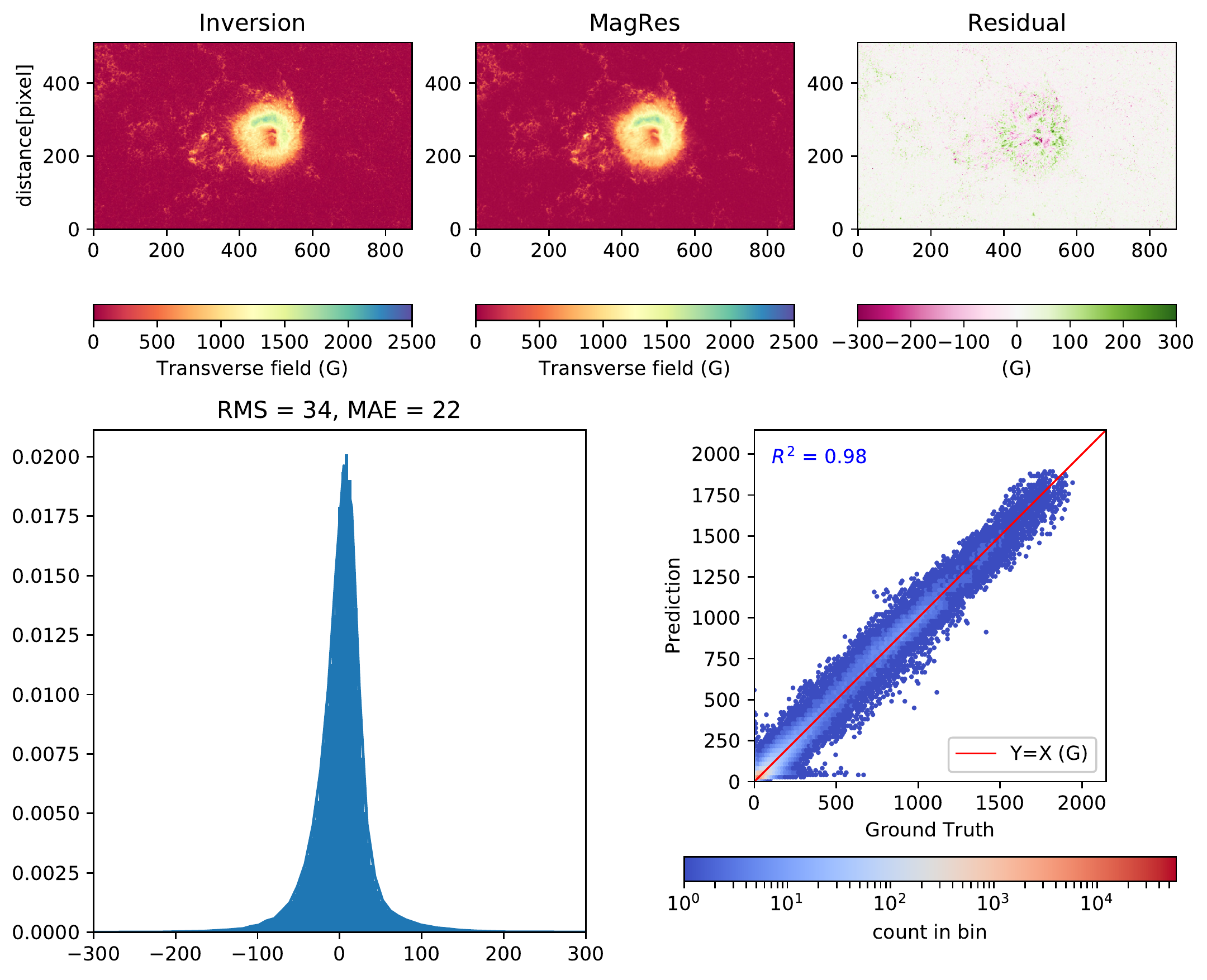}
	\caption{Upper left panel shows the inversion result for transverse magnetic flux density. The testing result is shown in the upper middle panel. The upper right panel is the residual error of the inversion result and the testing result. The lower left panel is the histogram of the residual error. The lower right panel is the scatter diagram, identifying the density of the inversion results with the testing results.}
\end{figure*}  

\section{Results and testing the network availability}
\label{sec:comp}
The MagReses are completely trained and finally have three convergent models to infer $ B_{l} $ (longitudinal magnetic flux density),  $ B_{t} $ (transverse magnetic flux density), and $ \varphi $ (azimuth angle). It takes 1.11s for the model to predict a map with a resolution of 384 pixel $ \times $ 684 pixel. The loss function $ MSE $ of a training set for $ B_{l} $ , $ B_{t} $, and $ \varphi $ models are 3522 G, 3313 G, and 525\degree\ (including noise). This means that the residual errors of training results with the target data for them are 59.3 G, 57.6 G, and 22.9$  \degree $. In our training process, the input images are set as any size. The data of test sets could be carried out to produce a physical quantity of a magnetic field without changing the size. Then Stokes $ I $, $ Q $, $ U $, and $ V $ only need to be normalized by Equation 6 before carrying out testing.

\begin{figure*}[!ht]
	\centering
	\includegraphics[width=1\linewidth]{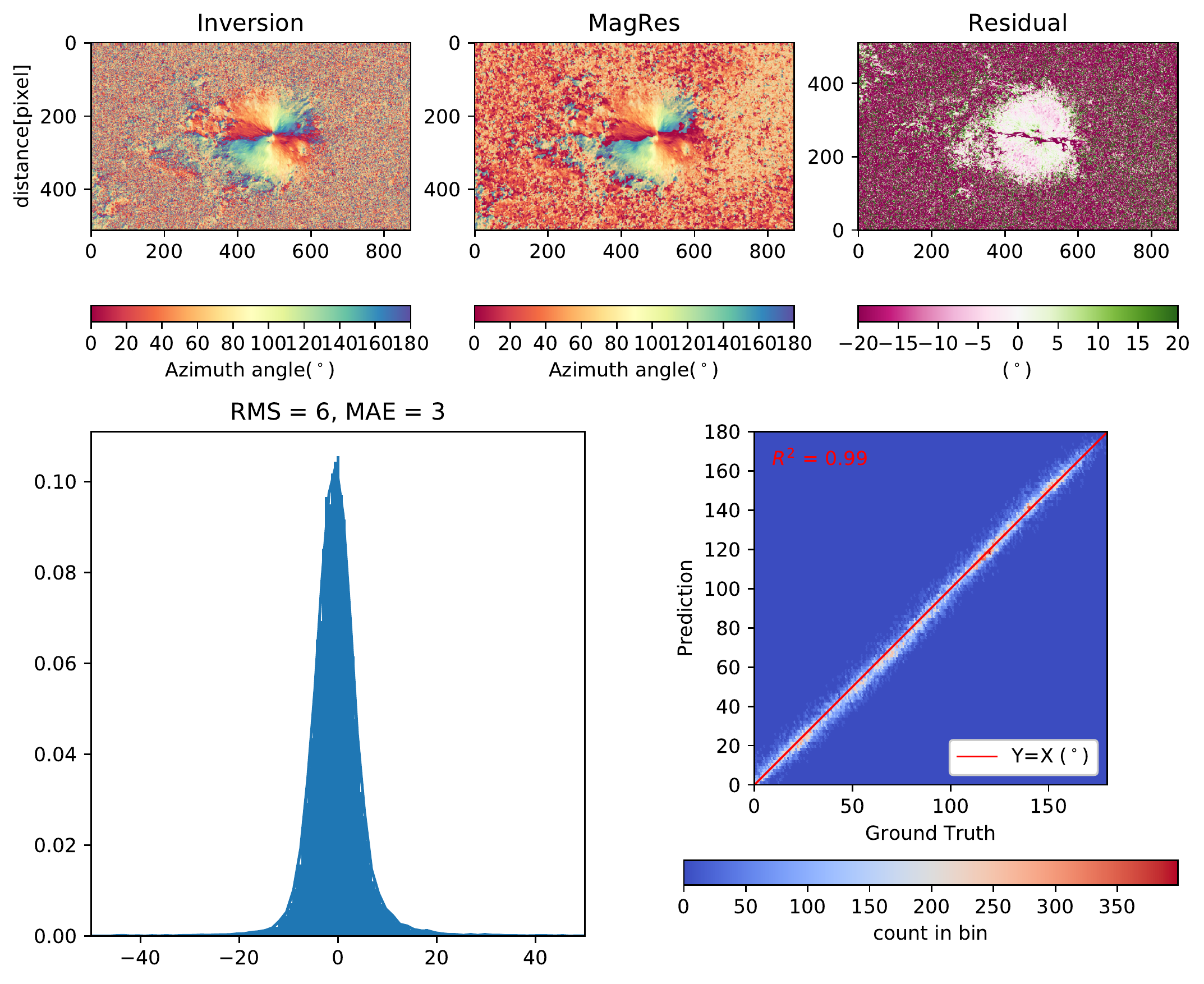}
	\caption{Upper left panel shows the inversion result for the azimuth angle. The testing result is shown in the upper middle panel. The upper right panel is the residual error of the inversion result and the testing result. The lower left panel is the histogram of the residual error. The lower right panel is the scatter diagram, identifying the density of the inversion results with the testing results.}
\end{figure*}

\begin{figure*}[h]
	\centering
	
	\begin{minipage}{1\linewidth}
		\centering
		\includegraphics[width=1\linewidth]{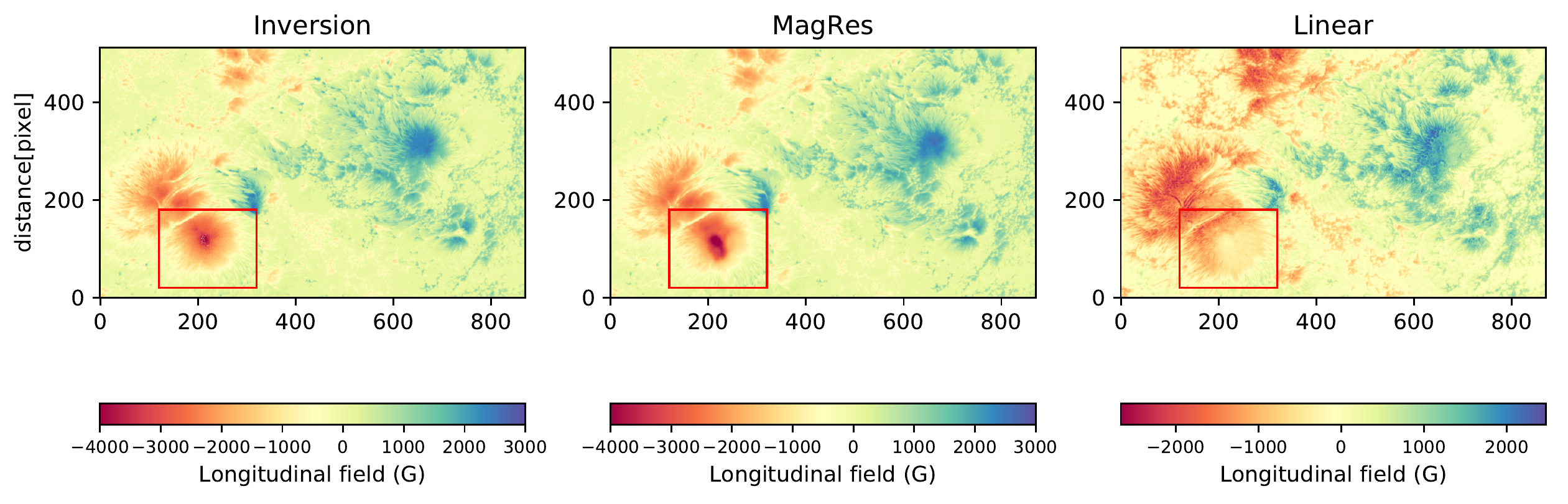}
	\end{minipage}

	\begin{minipage}{1\linewidth}
		\centering
		\includegraphics[width=1\linewidth]{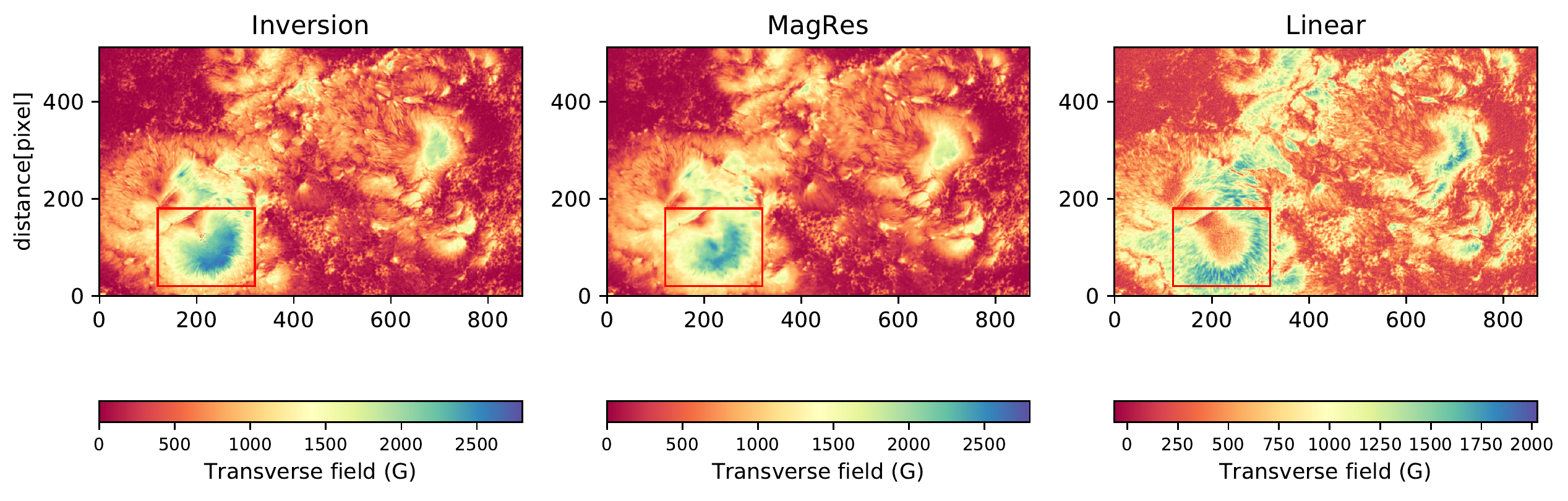}
	\end{minipage}%
	
	\caption{Results of the inversion, MagRes, and linear calibration for  $ B_{l} $ and $ B_{t} $. The top row is for $ B_{l} $, and the bottom row is for the $ B_{t} $.}
\end{figure*}

\begin{figure*}[h]
	\centering
	
	\begin{minipage}{1\linewidth}
		\centering
		\includegraphics[width=1\linewidth]{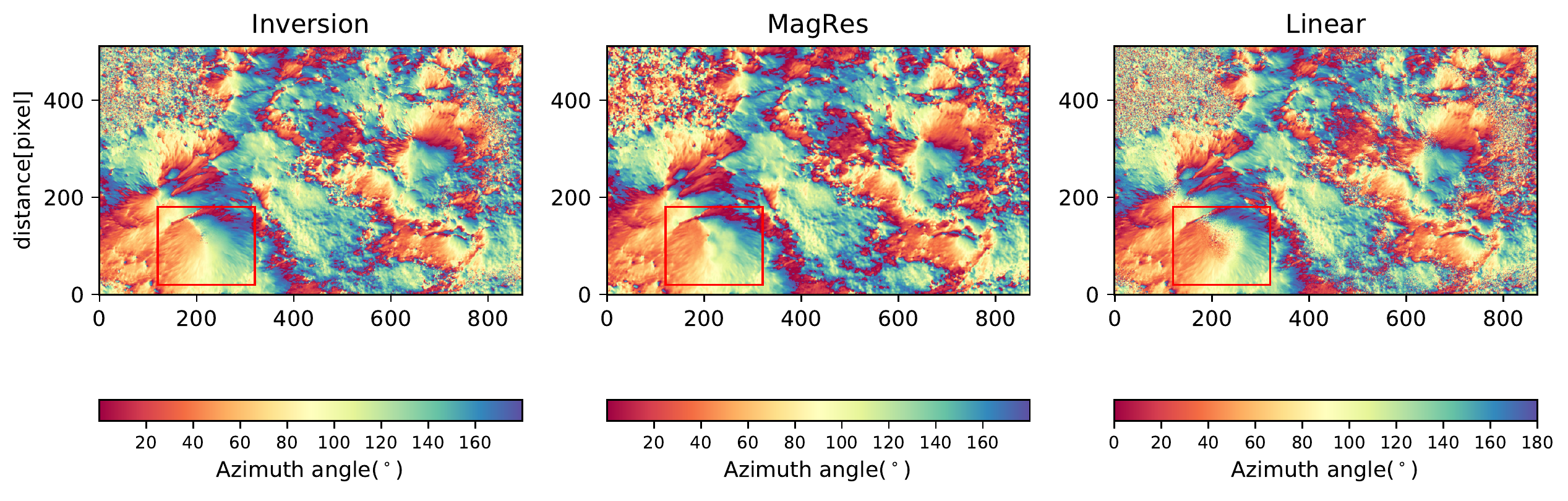}
	\end{minipage}

	\begin{minipage}{1\linewidth}
		\centering
		\includegraphics[width=1\linewidth]{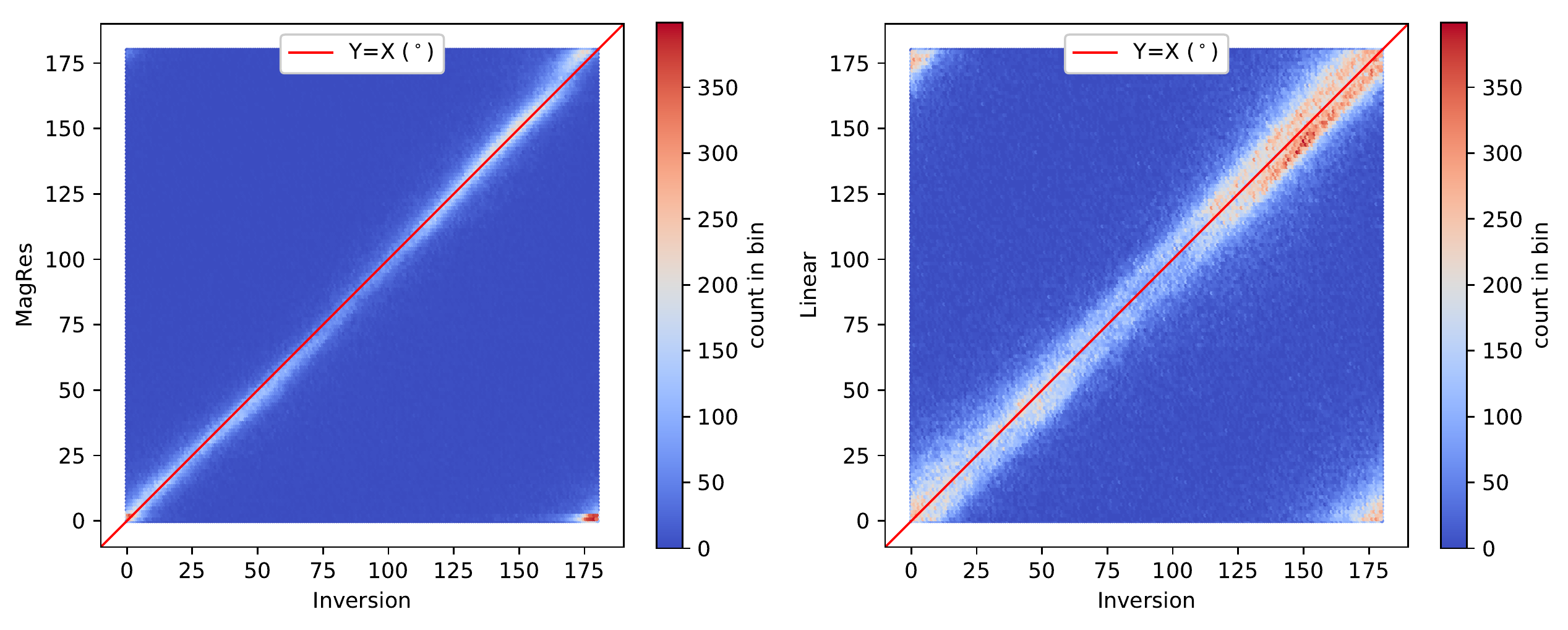}
	\end{minipage}%
	
	\caption{Results of the inversion, MagRes, and linear calibration for $ \varphi $. The scatter graph of the results of ResNet and the inversion method are displayed in the lower left. In lower right of the figure, the scatter graph of the results of     linear calibration and the inversion method can be seen.}
\end{figure*}

\subsection{Longitudinal magnetic flux density}

We employed Hinode/SP vector magnetic field data in NOAA AR 12738 as a test set to evaluate our model. The data were observed at 10:26 UT on 2019 April 13 with an image scale of 0.317 $ ^{\prime\prime} $ per pixel. Figure 4 presents the ResNet result for $B_{l}$. The ResNet $B_{l}$ is primarily similar to the inverted $B_{l}$, while it exhibits a cleaner appearance. This may be because the convolution operation smoothes the adjacent pixels and brings about de-noise in the maps. The residuals between ResNet $B_{l}$ and the inverted $B_{l}$ are mostly less than 300 G. In the sample, only 284 pixels in the residual map that have 446976 pixels are more than 300 G. The proportion of errors reaching 300 G in part of the active region is 0.002.

The lower left panel is the histogram of the residual error. We used the root mean square (RMS), which is defined as \begin{equation}
\mathrm{RMS} = \sqrt{\mathrm{MSE}},
\end{equation} of the residual errors as the effective value of the residual errors to evaluate the performance of the models. We used the mean absolute error (MAE), which is defined as \begin{equation}
\mathrm{MAE} = \dfrac{1}{n}\sum_{i=1}^{n}\lvert(y_{i}-\hat{y_{i}})\rvert
\end{equation}of the residual errors, as the mean value to evaluate the models. The RMS and MAE for the $ B_{l} $ are 35 G and 17 G. The ratio of MAE to RMS, which represents uncommonly large errors, is 0.48 here. The smaller this ratio is, the more unusual large residual errors are. The lower right panel is the scatter diagram, which identifies the density of the inversion results with the testing results. In statistics, the coefficient of determination, $ R^{2} $, is interpreted as the proportion of the variance in the dependent variable that is predictable from the independent variable, thus:\begin{equation}
R^{2} = 1- \dfrac{\sum_{i=1}^{n}(y_{i}-\hat{y_{i}})^{2}}{\sum_{i=1}^{n}(y_{i}-\overline{y})^{2}}.
\end{equation} The $ R^{2} $ represents for the accuracy of a fit. An $ R^{2} $ of 1 means the dependent variable can be predicted without error from the independent variable. In this case, $ R^{2} $ is 0.99.

\subsection{Transverse magnetic flux density}

Figure 5 presents the MagRes result for transverse magnetic flux density. The testing result of $ B_{t} $ in the upper middle panel has a very similar appearance as the inversion result in the upper left panel. The MagRes result in the upper middle panel is also cleaner than the inverted result. With the residual error, it is shown that the values of a few pixels exceed 300 G, which are dark purple or dark green; most of them are in the range of [-300 300] G. In the sample, only 85 pixels in the residual map are more than 300 G. The lower left panel is the histogram of the residual errors. The RMS and MAE of the residual errors are 34 G and 22 G. The MAE to the RMS ratio is 0.65 for transverse magnetic flux density, which means that there are fewer large, uncommon residual errors. The lower right panel is the scatter diagram, identifying the density of the inversion results with the testing results. The coefficient of determination $ R^{2} $ for the scatter diagram shown in the lower right panel is 0.98. 

\subsection{Azimuth angle}

The final results for $ \varphi $ are shown in Figure 6 and the contents are also the same as in Figure 4. There are many noises on the quiet region in the inversion results, as is shown in the upper left panel. In the sunspots region, MagRes calibratin has a similar appearance with the inversion results and it ignores the boundary area. The map of residual error in the upper right panel shows that most of the sunspot region is in the range of [-20 20] $ \degree $. We used the clean data that remove most of the quiet area by extracting the corresponding pixels, which are greater than 300G in transverse magnetic flux density and shown in lighter colors, to further analyze the performance below. The lower left panel is the histogram of the residual errors. The RMS and MAE of the residual errors are 6 $ \degree $ and 3 $ \degree $. The ratio of MAE to RMS is about 0.5. The lower right panel is the scatter diagram, which identifies the density of the inversion results with the testing results. The coefficient of determination $ R^{2} $ for the scatter diagram shown on the lower right panel is about 0.99. 

In conclusion, the new approach infers the magnetic field parameters with a precision comparable with that of the inversion technique. Furthermore, it produced cleaner maps with better noise suppression.

\section{Comparison and analysis of results}

\subsection{Comparison with the linear calibration on $ B_{l} $ and $ B_{t} $} 

One of our main aims for this study is to improve our model in regions with a strong magnetic field, where the linear calibration method fails for the magnetic saturation effect. Here we present the ResNets results in the magnetic saturation regions and conduct a comparison between the results of ResNets and the linear calibration method for $ B_{l} $ and $ B_{t} $ on the strong magnetic field.

The MagMLP tells us that the neural networks have the ability to solve the magnetic saturation. This is also the case for the CNN method, MagRes. We employ the active region AR 12192 data on the test set to demonstrate. It has a more complicated structure, which was observed with \textit{Hinode}/SP at 23:41 UT on 24 October 2014. Based on the weak field approximation, the linear calibrations for $ B_{l} $ and $ B_{t} $ are obtained by fitting the straight lines between the Stokes parameters and the inversion results using the least square method. The relationships for linear calibrations are presented in Equations 14 and 15:
\begin{equation}
B_{l}=2370.65 (V/I)-39.66,   
\end{equation}
\begin{equation}
B_{t}=2843.19 ((Q/I)^{2}+(U/I)^{2})^{1/4}-63.16.\end{equation} The results of MagRes, the linear calibration, and the Stokes inversion for $ B_{l} $ and $ B_{t} $ are shown in Figure 7. As is shown with red rectangles, the results of MagReses in the middle panels have a very similar appearance, structure, and shape as those from the inversion in the left panels. However, the results of linear calibration have pronounced magnetic saturation.

With the RMS of residual errors of target data with the testing results of networks, we compare and analyze the performance of the two different neural network methods. The RMS of residual errors for MagReses are less than half for those of MagMLP. The data from 2018 and 2019 in the test set act as samples, which can be seen in Table 2, indicating that the MagReses show a better performance than the MagMLPs based on the quantitative evaluation of the RMS. 

\begin{table*}[h]
	\caption{RMS values' residual errors of target data with the testing results of networks from 2018 and 2019 in the test set.}              
	\label{table:2}      
	\centering
	\begin{tabular}{cccccccc}
		\hline
		\multirow{2}{*}{\begin{tabular}[c]{@{}c@{}}Date\\ (yyyymmdd\_hhmmss)\end{tabular}}  & \multirow{2}{*}{Active region} & \multicolumn{2}{c}{Transverse field} & \multicolumn{2}{c}{Longitudinal field}  \\ \cline{7-8} 
		&                                & MagMLP            & MagRes           & MagMLP             & MagRes             \\ \hline
		20180206\_132552                      & AR 12699                       & 118               & 53               & 97                 & 44                \\
		20180210\_050505                      & AR 12699                       & 99                & 40               & 99                 & 49                \\
		20180212\_081503                      & AR 12699                       & 96                & 40               & 94                 & 47                \\
		20180422\_181519                      & AR 12703                       & 105               & 36               & 84                 & 39                \\
		20180621\_052804                      & AR 12715                       & 108               & 42               & 87                 & 45                \\
		20190321\_135705                      & AR 12736                       & 125               & 58               & 100                & 55                 \\
		20190411\_212122                      & AR 12738                       & 172               & 76               & 137                & 83                \\
		20190412\_180409                      & AR 12738                       & 151               & 61               & 122                & 68                \\
		20190413\_102645                      & AR 12738                       & 99                & 34               & 72                 & 35                \\
		20190416\_112936                      & AR 12738                       & 175               & 86               & 172                & 87                \\
		20190418\_072036                      & AR 12740                       & 163               & 88               & 206                & 100                \\
		20190508\_060010                      & AR 12740                       & 104               & 36               & 80                 & 38                \\ \hline
	\end{tabular}
\end{table*}
\begin{figure*}[!ht]
	\centering
	\includegraphics[width=1\linewidth]{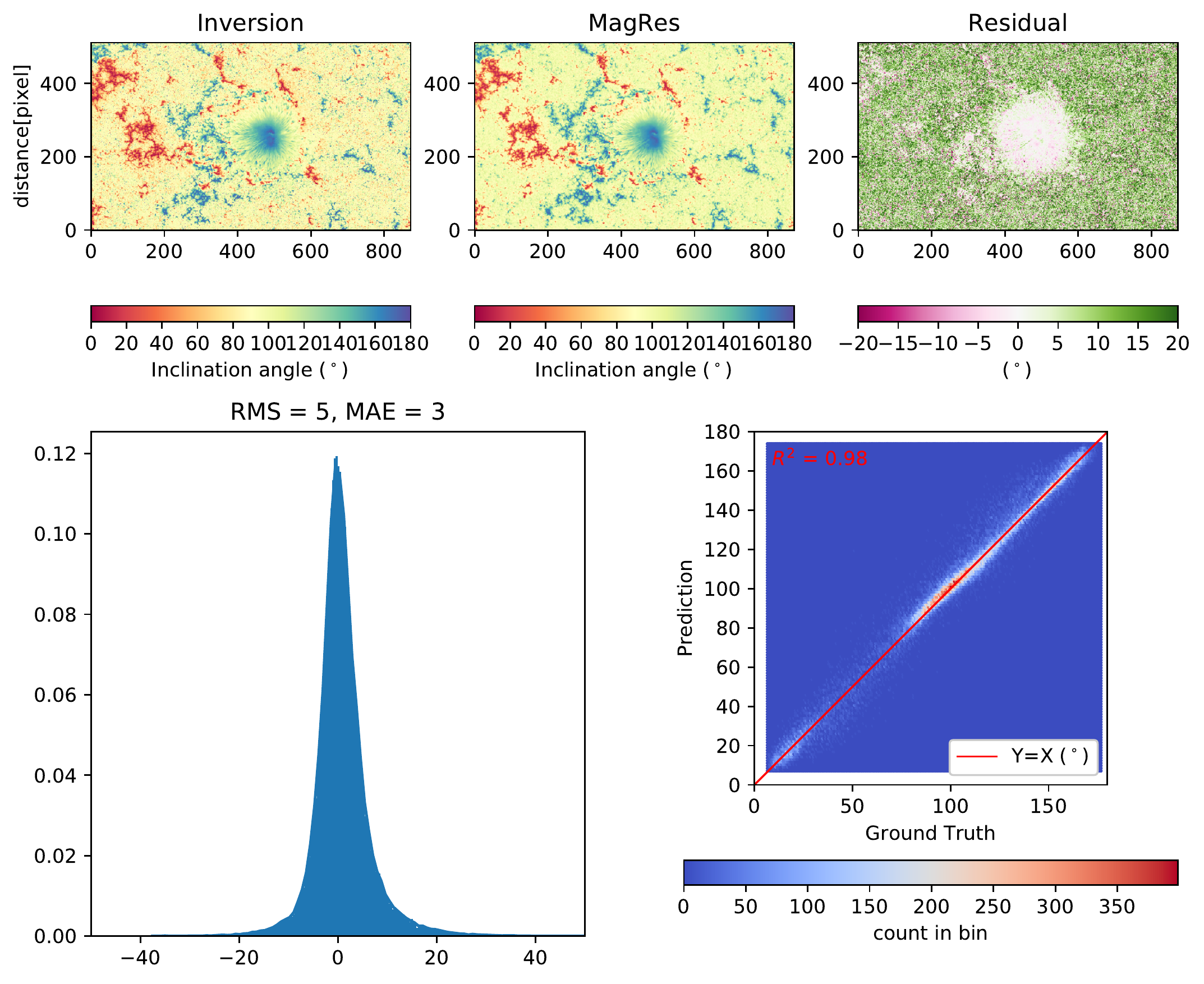}
	\caption{Upper left panel shows the inversion result for the inclination angle. The testing result is shown in the upper middle panel. The upper right panel is the residual error of the inversion result and the testing result. The lower left panel is the histogram of the residual error. The lower right panel is the scatter diagram, identifying the density of the inversion results with the testing results.}
\end{figure*}
\subsection{Comparison and analysis on results of azimuth angle}

The azimuth angle $ \varphi $ could be directly inferred by Equation 3. However, the effects of Faraday rotation exists in the filter-based magnetograph data, especially for the data taken near the line center \citep{2000SoPh..191..309H,2004SoPh..222...17S}. So we make a comparison for $ \varphi $ between the methods for ResNet and Equation 3. 

The $ \varphi $ (azimuth angle) is the direction of the magnetic field projected on the sky plane, and there is no magnetic saturation effect on $ \varphi $. Figure 8 compares the results of MagRes, the linear calibration, and Stokes inversions for $ \varphi $ in AR 12192. It displays that the results of ResNet and linear calibration are all similar to the inversion results. According to the scatter plots in the bottom panel of Figure 8, ResNet results are better approaching the inversion results than those of linear calibration. Additionally, the RMS of the residual error of the ResNet result with an inversion result is smaller than those for linear calibration, which are 16\degree and 21\degree, respectively. This indicates that ResNet could improve the accuracy of $ \varphi $.

\subsection{Analysis on the inclination angle}

Meanwhile, the $ \theta $ from the level 2 data of Hinode/SP have many ``bright spots'' and ``dark spots'' at 180 and 0 degrees respectively. In our previous paper \citep{Guo_2020}, the problem is discussed. We propose to do a comparison experiment with the previous work.

From Equations 4 and 5, we can infer that the $ \theta $ can be represented as:
\begin{equation}\theta=\arctan(\frac{B_{t}}{B_{l}}).\end{equation}
The results for $ \theta $ are shown in Figure 9. The testing result of $ \theta $ is similar to the inversion result in the sunspot region, but it presents significant differences in the quiet region. In our prior study, MagMLP models eliminate ``bright spots" and ``dark spots" from the inverted inclination angle map. However, these small-scale features are reconstructed by MagRes. The figure of the training result is cleaner. We extracted the corresponding pixels where the transverse magnetic flux density was above 150 G to further evaluate the model performance below that point. The map of residual error in the upper right panel illustrates that most of the active regions are in the range of [-20 20] $ \degree $ and there are many pixels in the quiet region outside of the range [-20 20] $ \degree $. The lower left panel is the histogram of the residual errors. The distribution is asymmetric, which may be because there are
more cases in which the number of pixels is higher than 90 $ \degree $ as opposed to lower than it. The RMS and MAE of the residual errors are 5 $ \degree $ and 3 $ \degree $. The ratio of MAE to RMS, which can help us understand whether there are large but uncommon errors, is about 0.6. This illustrates that there are a few uncommon residual errors which are larger. The lower right panel is the scatter diagram, which identifies the density of the inversion results with the testing results. The coefficient of determination $ R^{2} $ for the scatter diagram shown in the lower right panel is 0.98.

\section{Disscussion, conclusion, and future work}
\label{sec:conc}
We have developed a new approach for the magnetic field calibration for a filter-based magnetograph from a selected wavelength point of the Stokes profiles of \textit{Hinode}/SP based on ResNet. A series of experiments were carried out on $ B_{l} $ , $ B_{t} $, $ \varphi $, and $ \theta $. We collected 176 frames of data samples, including 121 frames for a training set, 18 frames for a validating set, and 37 frames for a test set. When training the network models, we considered the influential factors on the models' performance, such as data cleaning, data normalization, and $ \alpha $ (filling factor). The main effect of $ \alpha $ is in the quiet region. Considering it in the magnetic parameters will increase the prediction error of the models, especially in the weak magnetic field regions.

Different input parameters would generate results with a different accuracy. We have attempted to use Stokes $ V $ and $ I $ as input parameters to train the model of $ B_{l} $ and use Stokes $ Q $, $ U $, and $ I $ as input parameters to train the model of $ B_{t} $. We note that $ B_{l} $ and $ B_{t} $ can also be inferred from these models, but the accuracy is lower. Additionally, we attempted to select another wavelength point at this line profile to train the models. From the results of our experiments, these training results of models did not have a higher accuracy. The accuracy increased at a different degree, but these are just some of our attempts, and the research value for actual observations is not clear. We focused on the research of the wavelength position at -0.063 \AA\ from the center of line Fe \textsc{I} 6301 \AA\  since this is the closest to the routine observations of filter-based magnetograph.

Thereafter, the trained models were used to infer the vector magnetic fields from the samples of the test set. Firstly, the image data from AR 12738, which were collected on 13 April 2019, were utilized to test the models availability. Secondly, the image data from NOAA AR 12192, which were collected on 24 October 2014 with a more complicated magnetic structure, were used to display the comparison with the results of inversion and linear calibration. Thirdly, 12 frames of the image data from the test set were exploited to compare the performance of ResNet with MLP. Based on these experiments, we obtain the consistent finding, which can be summarized as follows:

(1) Our new method could infer the $ B_{l} $ (longitudinal magnetic flux density), $ B_{t} $ (transverse magnetic flux density), and $ \varphi $ (azimuth angle) well using narrow-band Stokes $ I $, $ Q $, and $ U $ as well as the $ V $ maps. The ResNet method produces cleaner magnetic maps with less noise compared with the inversion method. 

(2) The results inferred by the new approach are extremely close to the Stokes inversion results with the RMS values of residual errors within 100 G for  $ B_{l} $ and $ B_{t} $, where the RMS can be within 50 G for the simple AR images data. For $ \varphi $ (azimuth angle), the RMS reached 12\degree\ for the non-noisy simple structure data. The ResNet-inferred results are highly correlated to the inversion inferred results with the coefficient of determination $ R^{2} $ values being closer to 1.

(3) Compared with the linear calibration, our new approach could infer the magnetic fields better without the magnetic saturation.

(4) Based on the analysis of the RMS values of residual errors between the inversion inferred results and the MLP-produced or the ResNet-produced results, the ResNet-produced results are mostly below 50 G, which is much better than those of our previous multilayer perceptron (MLP) method, which were mostly above 100G. This may be because the new method could be able to utilize the spatial relationship between adjacent pixels on the input parameters.

The proposed method also has its shortcomings. Firstly, the final accuracy of the model is limited by the accuracy of the corresponding inversion method that the dataset used. Secondly, a large error occurs if the data predicted by using these models are beyond the data range of the training set. Thirdly, the S/N of actual observation data may also affect the recovered accuracy. All of these effects increase the difficulties involved in the selection of samples as well as cleaning and testing network models. One should keep the above effects in mind when using our method.

In conclusion, our attempts based on ResNet can be understood as an alternative, efficient solution to the problem of linear calibration for the filter-based magnetograph. The study is just a start, and more tests are needed to ensure that the magnetic field recovered from ResNet can be used for scientific analysis. We propose applying the ResNet method to FMG magnetic data calibration. Therefore, some experiments are being conducted on the full-disk data using the observation data of \textit{Helioseismic and Magnetic Imager} onboard the \textit{Solar Dynamics Observatory} \citep{2012SoPh..275..229S}, considering the influence of the orbital velocity. Meanwhile, we are also trying to use other neural networks to train models, such as the Dense Convolutional Network (DenseNet) \citep{2016arXiv160806993H}. In addition, neural networks are not well designed to work with circular quantities, such as the azimuth angle. Whether the function approximation conditions can be considered based on these physical requirements (such as integrating physical constraints on the basis of the traditional loss function) is also a question worth studying. This is not only an important issue for this project, it will affect the degree of trust and use of these data processing results by solar physicists. We also want to use this research to explore a way to overcome the application of machine learning methods in astronomy methods of cognitive impairment to better serve astronomy research in the future.

\acknowledgements
 We are very grateful to our referee for putting forward many valuable feedback. We thank the Astronomical Big Data Joint Research Center, co-founded by the National Astronomical Observatories, the Chinese Academy of Sciences and the Alibaba Cloud. Thanks for the Community Spectropolarimetric Analysis Center (\url{http://www2.hao.ucar.edu/csac}) providing Hinode/SP data. This project has received funding from the Strategic Priority Research Program on Space Science, the Chinese Academy of Sciences under No. XDA15320300, XDA15320302, XDA15052200, XDA15010800, the National Natural Science Foundation of China (NSFC) under No.12073077, 11873027, 11773072, 11427803, 11427901, 11773040, 11573012, 11833010, 11973056, 11873062, 11773038, 11703042, and Beijing Municipal Science and Technology under No. Z181100002918004. We also thank the NVIDIA Corporation for the donation of the Quadro P2000, one of the GPUs in this work. We acknowledge the community effort devoted to the development of the following open-source packages used in the research: Keras (keras.io), TensorFlow (tensorflow.org), Matplotlib (matplotlib.org), Numpy (numpy.org), and Astropy (astropy.org). We are grateful to prof. Hongqi Zhang and Dr. Junfeng Hou of HSOS for helpful discussion.

\bibliographystyle{aasjournal}
\bibliography{ref}

\appendix
\section*{appendix}
\begin{figure*}[!ht]
	\centering
	\includegraphics[width=1\linewidth]{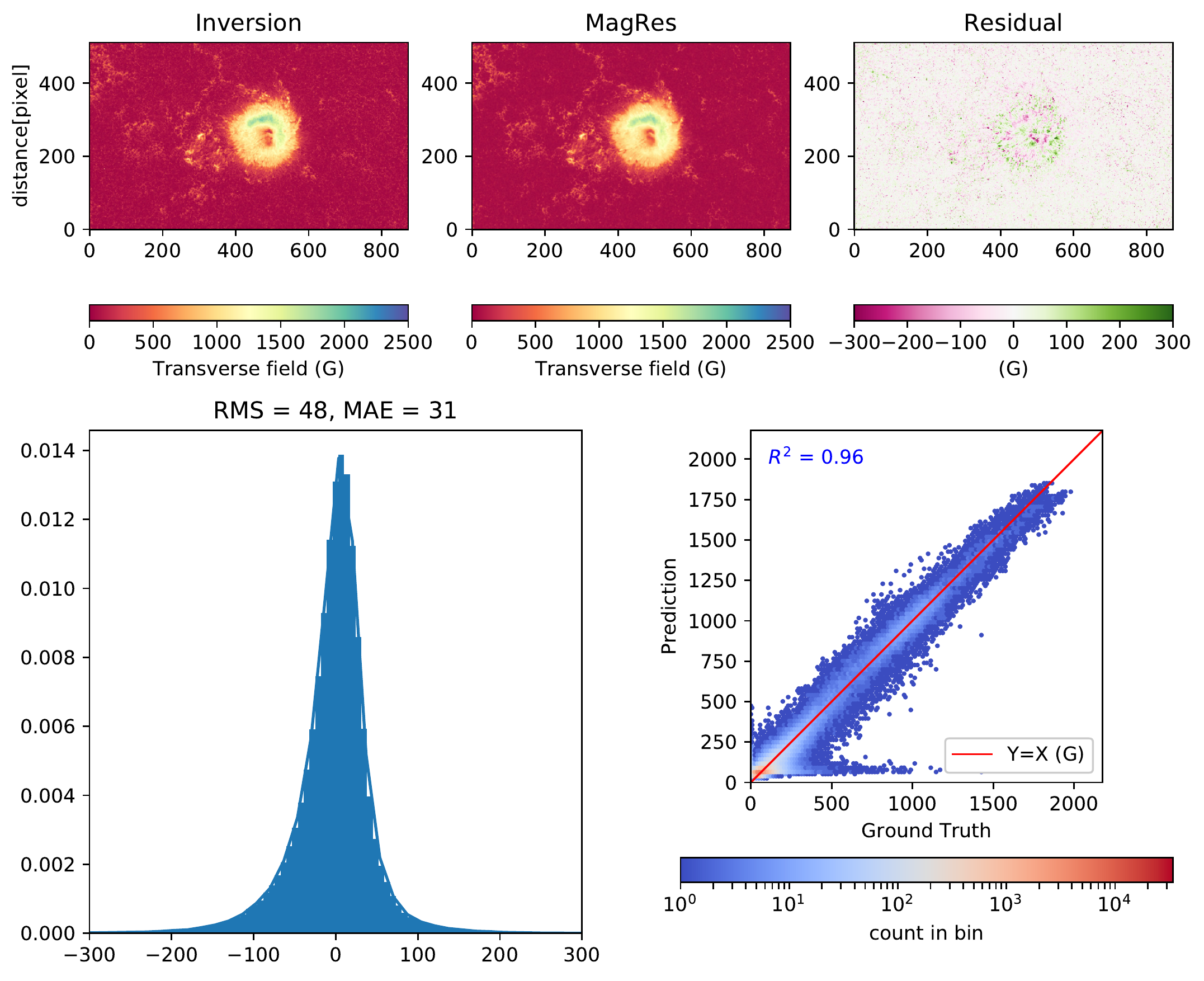}
	\caption{Upper left panel shows the inversion result for  $ B_{tf} $. The testing result is shown in the upper middle panel. The upper right panel is the residual error of the inversion result and the testing result. The lower left panel is the histogram of the residual error. The lower right panel is the scatter diagram, identifying the density of the inversion results with the testing results.}
\end{figure*}  
\subsection{Transverse magnetic flux density and filling factor}
As previously mentioned footnote 1, the observed Stokes \textit{I} profile $ I_{obs} $ is fitted with $ \alpha I_{mag} +(1-\alpha ) I_{scatt} $, indicating that $ Q=\alpha Q_{mag} $, $ U=\alpha U_{mag} $, and $ V=\alpha V_{mag} $. In the weak field regime, Stokes $ Q $ and $ U $ are proportional to the squared transverse magnetic flux density \citep{2019LRSP...16....1B}. In other words, $ B_{tf} $, without considering the $ \alpha $, follow Equation .17 based on the weak field regime:
\begin{equation}B_{tf}=\sqrt\alpha B \sin(\theta ).\end{equation}
This is because alpha acts in the signals and not in the magnetic field itself. If an approximation like weak field is used, the derived quantity has an impact on the output, that is, the magnetic field inferred does not generate the observed signals in areas with a small filling factor and a large field. It could work inside the umbra where alpha is high and the field is large, but not in a plage or faculae where the field is high and the filling factor is small (due to the saturation effect). If one follows the theory and Equation 16, one cannot use Equation 15 to calculate the inclination and one would need the information of alpha to do that.

We also trained a ResNet for $ B_{tf} $; the results of which can be seen in Figure 1. The results of ResNet appear to be consistent with the inversion results. The RMS and MAE of the residual errors are 48 G and 31 G. The coefficient of determination $ R^{2} $ for the scatter diagram shown in the lower right panel is 0.96. These indicators show poorer results than $ B_{t} $. On the quiet region, we can see that more dots are distributed in the upper right panel than those of $ B_{t} $, which are shown in Figure 5.

\end{document}